\documentstyle[aps,pre,preprint,epsf]{revtex}
\def\LL  {\Lambda ^{\!\!\scriptscriptstyle-1}}
\def\B.#1{{\bbox{#1}}}
\def\BC.#1{{\bbox{\cal{#1}}}}

\begin{document}
\title{Correlation functions in isotropic and anisotropic turbulence: the role of
the symmetry group}
\author{Itai Arad, Victor S. L'vov and Itamar Procaccia}
\address{Department of ~~Chemical Physics, The Weizmann Institute of Science\\Rehovot,
76100, Israel}
\maketitle

\begin{abstract}
The theory of fully developed turbulence is usually considered in an
idealized homogeneous and isotropic state. Real turbulent flows exhibit the
effects of anisotropic forcing. The analysis of correlation functions and
structure functions in isotropic and anisotropic situations is facilitated
and made rational when performed in terms of the irreducible representations
of the relevant symmetry group which is the group of all rotations SO(3). In
this paper we firstly consider the needed general theory and explain why
we expect different (universal) scaling exponents in the different sectors
of the symmetry group. We exemplify the theory 
context of isotropic turbulence (for third order tensorial structure functions)
and in weakly anisotropic turbulence (for the second order structure function). The
utility of the resulting expressions for the analysis of experimental data
is demonstrated in the context of high Reynolds number measurements of
turbulence in the atmosphere.
\end{abstract}


\section{Introduction}

Experiments in fluid turbulence are usually limited to the measurement of
the velocity field at one single spatial point as a function of time. This
situation has begun to improve recently, but still much of the analysis of
the statistical properties of Navier-Stokes turbulence \cite{Fri} is
influenced by this tradition: the Taylor hypothesis is used to justify the
identification of velocity signals at different times with differences of
longitudinal velocity components across a spatial length scale $R$. Most of
the available statistical information is therefore about properties of
longitudinal two-point differences of the Eulerian velocity field and their
moments, termed structure functions: 
\begin{equation}
S_{n}(R)=\langle |[{\B.u}({\B.r}+{\B.R})-{\B.u}({\B.r})]\cdot \frac{{\B.R}}{R
}|^{n}\rangle \ ,  \label{sSn}
\end{equation}
where $\langle \cdot \rangle $ denotes ensemble averaging. In isotropic
homogeneous turbulence, these structure functions are expected to behave as
a power-law in $R$, $S_{n}(R)\sim R^{\zeta _{n}}$, with apparently universal
scaling exponents $\zeta _{n}$.

Recent research \cite{96LP-2,99BLPP} has pointed out the advantages of considering not only the
longitudinal structure functions, but tensorial multi-point correlations of
velocity field differences 
\begin{equation}
{\B.w}({\B.r},{\B.r}^{\prime },t)\equiv {\B.u}({\B.r}^{\prime },t)-{\B.u}({
\B.r},t),  \label{du1}
\end{equation}
given by 
\begin{eqnarray}
&&\hat{{\cal F}}^{\alpha\beta\dots\delta}_{n}({\B.r}_{1},{\B.r}_{1}^{\prime },t_{1};{\B.r}_{2},{
\B.r}_{2}^{\prime },t_{2};\dots ;{\B.r}_{n},{\B.r}_{n}^{\prime },t_{n}) 
\nonumber \\
&=&\langle w^\alpha({\B.r}_{1},{\B.r}_{1}^{\prime },t_{1})w^\beta({\B.r}_{2},{
\B.r}_{2}^{\prime },t_{2})\dots w^\delta({\B.r}_{n},{\B.r}_{n}^{\prime
},t_{n})\rangle \ ,  \label{defFtilde}
\end{eqnarray}
where all the coordinates are distinct. Even when the coordinates fuse to
yield time-independent structure functions depending on one separation only,
these are tensorial quantities \cite{97LPP} denoted as 
\begin{equation}
S^{\alpha \beta \dots }({\B.R})\equiv \langle \lbrack u^{\alpha }({\B.r}+{\B.
R})-u^{\alpha }({\B.r})][u^{\beta }({\B.r}+{\B.R})-u^{\beta }({\B.r})]\dots
\rangle \ .  \label{Sn}
\end{equation}
Needless to say, the tensorial information is lost in the usual measurements
leading to (\ref{sSn}). One of the main points of the present paper is that
keeping the tensorial information can help significantly in disentangling
different scaling contributions to the statistical objects, contributions
that are hard to distinguish when quantities like (\ref{sSn}) are
considered. Especially when anisotropy implies different tensorial
components with possible different scaling exponents characterizing them,
careful control of the various contributions is called for.

To understand why irreducible representations of the symmetry group may have
an important role in determining the form of correlation functions, we need
to discuss the equations of motion which they satisfy. We shall show that
the isotropy of the Navier-Stokes equation and the incompressibility
condition implies the isotropy of the hierarchical equations which the
correlation functions satisfy. We will use this symmetry to show that every
component of the general solution with a definite behavior under rotations
(i.e., components of a definite {\em irreducible representation} of the $
SO(3)$ group) has to satisfy these equations by itself - independently of
components with different behavior under rotations. This ``foliation'' of
the hierarchical equations may possibly lead to different scaling exponents for
each component of the general solution which belong to a different $SO(3)$
irreducible representation.

In Sect.2 we describe the general mathematical framework of theory by
discussing the structure of tensorial fields from the point of view of $
SO(3) $ irreducible representations. We then show in Sect.3 that the
hierarchy equations are indeed isotropic and as a result foliate into
different sectors of the $SO(3)$ irreducible representations. In the next
sections we demonstrate the utility of the theory. In Sect.4 we revisit
Kolmogorov's four fifth's law emphasizing the rule of the $SO(3)$
irreducible representations in its derivation. Then, in Sect.5, we present
some experimental evidences for the importance of an anisotropic exponent in
the second order structure function, in atmospheric measurements. Sect.6
offers conclusions and some comments about the road ahead. 

\section{Tensorial correlation functions and $SO(3)$ irreducible
representations: general theory}

The physical objects that we deal with are the moments of the velocity
field at different space-time locations. In this section we suggest a way of
decomposing these objects into components with a definite behavior under
rotations. We will show later that components with different behavior under
rotation are subject to different dynamical equations, and therefore,
possibly, scale differently. Essentially, we are about to describe the
tensorial generalization of the well-known procedure of decomposing a scalar
function $\Psi ({\bf r})$ into components of different irreducible
representations using the spherical harmonics: 
\begin{equation}
\Psi ({\bf r})=\sum_{l,m}a_{lm}(r)Y_{lm}(\hat{{\bf r}})\ .
\label{scalar-case}
\end{equation}

\subsection{Formal definition}

Consider a typical moment of the velocity field, Eq.(\ref{defFtilde}). 
${ F}_{n}^{\alpha _{1}\ldots \alpha _{n}}({\bf r}_{1},{\bf r}
_{1}^{\prime },t_{1};\ldots ;{\bf r}_{n},{\bf r}_{n}^{\prime },t_{n})$ is a
function of $2n$ spatial variables and $n$ temporal variables. Physically,
it is a {\em tensor field}: if ${\bf F}_n$ is measured in two frames $I$
and $\overline{I}$ which are connected by the spatial transformation (say, a
rotation) 
\begin{equation}
\overline{x}^{\alpha }=\Lambda ^{\alpha }{}_{\beta }x^{\beta }
\label{coordinate-rotation}
\end{equation}
then, the measured quantities in each frame will be connected by the
relation: 
\begin{eqnarray}
&&\overline{{ F}}_{n}^{\alpha _{1}\ldots \alpha _{n}}(\overline{{\bf r}}
_{1},\overline{{\bf r}}_{1}^{\prime },\overline{t}_{1};\ldots ;\overline{
{\bf r}}_{n},\overline{{\bf r}}_{n}^{\prime },\overline{t}_{n})
\label{tensor-trans} \\
&=&\Lambda ^{\alpha _{1}}{}_{\beta _{1}}\cdots \Lambda ^{\alpha
_{n}}{}_{\beta _{n}}{ F}_{n}^{\beta _{1}\ldots \beta _{n}}({\bf r}_{1},
{\bf r}_{1}^{\prime },t_{1};\ldots ;{\bf r}_{n},{\bf r}_{n}^{\prime },t_{n})
\nonumber \\
&=&\Lambda ^{\alpha _{1}}{}_{\beta _{1}}\cdots \Lambda ^{\alpha
_{n}}{}_{\beta _{n}}{ F}_{n}^{\beta _{1}\ldots \beta _{n}}(\LL\overline{
{\bf r}}_{1},\LL\overline{{\bf r}}_{1}^{\prime },\overline{t}_{1};\ldots ;\LL
\overline{{\bf r}}_{n},\LL\overline{{\bf r}}_{n}^{\prime },\overline{t}
_{n})\ .  \nonumber
\end{eqnarray}
We see that as we move from one frame to another, the {\em functional form}
of the tensor field changes. We want to classify the different tensor fields
according to the change in their functional form as we make that move. We
can omit the time variables from our discussion since under rotation they
merely serve as parameters.

Consider coordinate transformations which are pure rotations. For such
transformations we may simplify the discussion by separating the dependence
on the amplitude of ${\bf r}_{i}$ from the dependence on the directionality
of ${\bf r}_{i}$: 
\begin{eqnarray*}
&&T^{\alpha _{1}\ldots \alpha _{n}}({\bf r}_{1},\ldots ,{\bf r}_{p}) \\
&=&T^{\alpha _{1}\ldots \alpha _{n}}(r_{1},\ldots ,r_{p};\hat{{\bf r}}
_{1},\ldots ,\hat{{\bf r}}_{p})
\end{eqnarray*}
For pure rotations we may treat the amplitudes $r_{1},\ldots ,r_{p}$ as
parameters: the transformations properties of $T^{\alpha _{1}\ldots \alpha
_{n}}$ under rotation are determined only by the dependence of $T^{\alpha
_{1}\ldots \alpha _{n}}$ on the unit vectors $\hat{{\bf r}}_{1},\ldots ,\hat{
{\bf r}} _{p}$. Accordingly it seems worthwhile to discuss tensor fields
which are functions of the unit vectors {\em only}. Notice that in the
scalar case we follow the same procedure: by restricting our attention to
scalar functions that depend only on the unit vector $\hat{{\bf r}}$, we
construct the spherical harmonics. These functions are {\em defined} such
that each one of them has unique\ transformation properties under rotations.
We then represent the most general scalar function as a linear combination
of the spherical harmonics with $r$-dependent coefficients, see Eq. (\ref
{scalar-case}).

The classification of the tensor fields $T^{\alpha _{1}\ldots \alpha _{n}}( 
\hat{{\bf r}}_{1},\ldots ,\hat{{\bf r}}_{p})$ according to their functional
change under rotations follows immediately from group representation theory
\cite{Corn,Stern}.
But in order to demonstrate that, we must first make some formal
definitions. We define ${\cal S}_{p}^{n}$ to be the space of all smooth
tensor fields of rank $n$ which depend on $p$ unit vectors. This is
obviously a linear space of infinite dimension. With each rotation $\Lambda
\in SO(3)$, we may now associate a linear transformation ${\cal O}_{\Lambda }$ 
on that space via the relation (\ref{tensor-trans}): 
\begin{eqnarray*}
&&\left[ {\cal O}_{\Lambda }T\right] ^{\alpha _{1},\ldots ,\alpha _{n}}(\hat{
{\bf r}}_{1},\ldots ,\hat{{\bf r}}_{p}) \\
&\equiv &\Lambda ^{\alpha _{1}}{}_{\beta _{1}}\cdots \Lambda ^{\alpha
_{n}}{}_{\beta _{n}}T^{\beta _{1}\ldots \beta _{n}}(\LL\hat{{\bf r}}
_{1},\ldots ,\LL\hat{{\bf r}}_{p}).
\end{eqnarray*}
Using this definition, it is easy to see that the set of linear operators $
{\cal O}_{\Lambda }$ furnishes a representation of the rotation group $SO(3)$
since they satisfy the relations: 
\begin{eqnarray*}
{\cal O}_{\Lambda _{1}}{\cal O}_{\Lambda _{2}} &=&{\cal O}_{\Lambda
_{1}\Lambda _{2}} \\
{\cal O}_{\Lambda }^{-1} &=&{\cal O}_{\Lambda ^{-1}}.
\end{eqnarray*}
General group theoretical considerations imply that it is possible to
decompose ${\cal S}_{p}^{n}$ into subspaces which are invariant to the
action of all the group operators ${\cal O}_{\Lambda }$. Moreover, we can
choose these subspaces to be {\em irreducible} in the sense that they will
not contain any invariant subspace themselves (excluding themselves and the
trivial subspace of the zero tensor field). For the $SO(3)$ group each of
these subspaces is conventionally characterized by an integer $
j=0,1,2,\ldots $ and is of dimension $2j+1$ \cite{Corn,Stern}. It should be noted that unlike
the scalar case, in the general space ${\cal S}_{p}^{n}$, there might be
more than one subspace for each given value of $j$. We therefore use the
index $q$ to distinguish subspaces with the same $j$. For each irreducible
subspace $(q,j)$ we can now choose a basis with $2j+1$ components labeled by
the index $m$: 
\[
B_{qjm}^{\alpha _{1},\ldots ,\alpha _{n}}(\hat{{\bf r}}_{1},\ldots ,\hat{
{\bf r}}_{p})\;;\;m=-j,\ldots +j. 
\]
In each subspace $(q,j)$, the group operators ${\cal O}_{\Lambda }$ furnish
a $2j+1$ dimensional irreducible representation of $SO(3)$. Using the basis $
B_{qjm}^{\alpha _{1},\ldots ,\alpha _{n}}(\hat{{\bf r}}_{1},\ldots ,\hat{
{\bf r}}_{p})$, we can represent each operator ${\cal O}_{\Lambda }$ as a $
(2j+1)\times (2j+1)$ matrix $D_{m^{\prime }m}^{(j)}(\Lambda )$ via the
relation: 
\begin{eqnarray*}
&&\left[ {\cal O}_{\Lambda }B\right] _{qjm}^{\alpha _{1},\ldots ,\alpha
_{n}}(\hat{{\bf r}}_{1},\ldots ,\hat{{\bf r}}_{p}) \\
&=&\Lambda ^{\alpha _{1}}{}_{\beta _{1}}\cdots \Lambda ^{\alpha
_{n}}{}_{\beta _{n}}B _{qjm}^{\beta _{1}\ldots \beta _{n}}(\LL\hat{{\bf r}}
_{1},\ldots ,\LL\hat{{\bf r}}_{p}) \\
&\equiv &\sum_{m^{\prime }=-j}^{+j}D_{m^{\prime }m}^{(j)}(\Lambda
)B_{qjm^{\prime }}^{\alpha _{1},\ldots ,\alpha _{n}}(\hat{{\bf r}}
_{1},\ldots ,\hat{{\bf r}}_{p}).
\end{eqnarray*}
It is conventional to choose the basis ${\bf B}_{qjm}$ such that the
matrices $D_{m^{\prime }m}^{(j)}(\phi )$, that correspond to rotations of $
\phi $ radians around the 3 axis, will be diagonal, and given by: 
\[
D_{m^{\prime }m}^{(j)}(\phi )=\delta _{mm^{\prime }}e^{im\phi }. 
\]

The ${\cal S}_{p}^{n}$ space possesses a very natural inner-product: 
\begin{eqnarray*}
\left\langle {\bf T},{\bf U}\right\rangle \!\! &\equiv &\!\!\int \!\!d\hat{
{\bf x}}_{1}\dots d\hat{{\bf x}}_{p} \\
&&\cdot T^{\alpha _{1}\dots \alpha _{n}}(\hat{{\bf x}}_{1}\dots \hat{{\bf x}}
_{p})g_{\alpha _{1}\beta _{1}}\ldots g_{\alpha _{n}\beta _{n}}U^{^{\beta
_{1}\ldots \beta _{n}}}(\hat{{\bf x}}_{1}\dots \hat{{\bf x}}_{p})^{\ast }\ 
\end{eqnarray*}
where $g_{\alpha \beta }$ is the 3-dimensional Euclidean metric tensor: 
\[
g_{\alpha \beta }=\left( 
\begin{array}{ccc}
1 &  &  \\ 
& 1 &  \\ 
&  & 1
\end{array}
\right) . 
\]
By definition, the rotation matrices $\Lambda ^{\alpha }{}_{\beta }$
preserve this metric, and therefore it is easy to see that for each $\Lambda
\in SO(3)$ we get: 
\[
\left\langle {\cal O}_{\Lambda }{\bf T},{\cal O}_{\Lambda }{\bf U}
\right\rangle =\left\langle {\bf T},{\bf U}\right\rangle 
\]
so that, ${\cal O}_{\Lambda }$ are unitary operators. If we now choose the
basis ${\bf B}_{qjm}$ to be orthonormal with respect to the inner-product
defined above, then the matrices $D_{m^{\prime }m}^{(j)}(\Lambda )$ will be
unitary.

Finally, consider {\em isotropic tensor fields}. An isotropic tensor field
is a tensor field which preserves its functional form under any arbitrary
rotation of the coordinate system. In other words, it is a tensor field
which is invariant to the action of all operators ${\cal O}_{\Lambda }$. The
one dimensional subspace spanned by this tensor-field is therefore invariant
under all operators ${\cal O} _{\Lambda }$ and therefore it must be a $j=0$
subspace.

Once the basis ${\bf B}_{qjm}$ has been selected, we may expand any
arbitrary tensor field $F^{\alpha _{1}\ldots \alpha _{n}}({\bf r}_{1},\ldots
,{\bf r}_{p})$ in this basis. As mentioned above, for each fixed set of
amplitudes $r_{1},\ldots ,r_{p}$, we can regard the tensor field $F^{\alpha
_{1}\ldots \alpha _{n}}({\bf r}_{1},\ldots ,{\bf r}_{p})$ as a tensor field
which depends only on the unit vectors $\hat{{\bf r}}_{1},\ldots ,\hat{{\bf 
r }}_{p}$, and hence belongs to ${\cal S}_{p}^{n}$. We can therefore expand
it in terms of the basis tensor fields ${\bf B}_{qjm}$ with coefficients
that depend on $r_{1},\ldots ,r_{p}$: 
\begin{eqnarray}
&&F^{\alpha _{1}\ldots \alpha _{n}}({\bf r}_{1},\ldots ,{\bf r}_{p}) 
\nonumber \\
&=&\sum_{q,j,m}a_{qjm}(r_{1},\ldots ,r_{p})B_{qjm}^{\alpha _{1},\ldots
,\alpha _{n}}(\hat{{\bf r}}_{1},\ldots ,\hat{{\bf r}}_{p}) \ .  \label{Texp}
\end{eqnarray}
The goal of the following sections is to demonstrate the utility of such
expansions for the study of the scaling properties of the correlation
functions.


\subsection{Construction of the basis tensors}

\paragraph{The Clebsch-Gordan machinery}.

A straightforward (although somewhat impractical) way to construct the basis
tensors ${\bf B}_{qjm}$ is to use the well-known Clebsch-Gordan machinery.
In this approach we consider the ${\cal S}_{p}^{n}$ space as a {\em direct
product space} of $n$ 3-dimensional Euclidean vector spaces with $p$
infinite dimensional spaces of single-variable continuous functions on the
unit sphere. In other words, we notice that ${\cal S}_{p}^{n}$ is given by: 
\[
{\cal S}_{p}^{n}=\left[ {\cal S}_{0}^{1}\right] ^{n}\otimes \left[ {\cal S}
_{1}^{0}\right] ^{p}, 
\]
and therefore every tensor $T^{\alpha _{1}\dots \alpha _{n}}(\hat{{\bf r}}
_{1}\dots \hat{{\bf r}}_{p})$ can be represented as a linear combination of
tensors of the form: 
\[
v_{1}^{\alpha _{1}}\ldots v_{n}^{\alpha _{n}}\varphi _{{\small 1}}\left( 
\hat{{\bf r}}_{1}\right) \cdot \ldots \cdot \varphi _{p}\left( \hat{{\bf r}}
_{p}\right) . 
\]
$v_{i}^{\alpha _{i}}$ are constant Euclidean vectors and $\varphi _{i}(\hat{
{\bf r}}_{i})$ are continuous functions over the unit sphere. The
3-dimensional Euclidean vector space, ${\cal S}_{0}^{1}$, contains exactly
one irreducible representation of $SO(3)$ - the $j=1$ representation - while 
${\cal S}_{1}^{0}$, the space of continuous functions over the unit sphere,
contains every irreducible representation exactly once. The statement that $
{\cal S}_{p}^{n}$ is a direct product space may now be written in a group
representation notation as: 
\[
{\cal S}_{p}^{n}=\stackrel{n\text{ \ times}}{\overbrace{1\otimes 1\otimes
\ldots \otimes 1}}\otimes \stackrel{p\text{ \ times}}{\overbrace{\left(
0\oplus 1\oplus 2\dots \right) \otimes \ldots \left( 0\oplus 1\oplus 2\dots
\right) }} 
\]

We can now choose an appropriate basis for each space in the product:

\begin{itemize}
\item  For the 3-dimensional Euclidean space we may choose: 
\[
{\bf e}_{1}={\frac{1}{\sqrt{2}}}\left( 
\begin{array}{l}
1 \\ 
i \\ 
0
\end{array}
\right) ,\quad {\bf e}_{0}={\frac{1}{\sqrt{2}}}\left( 
\begin{array}{l}
0 \\ 
0 \\ 
1
\end{array}
\right) ,\quad {\bf e}_{-1}={\frac{1}{\sqrt{2}}}\left( 
\begin{array}{l}
1 \\ 
-i \\ 
0
\end{array}
\right) 
\]

\item  For the space of continuous functions over the unit sphere we may
choose the well-known spherical harmonic functions.
\end{itemize}

Once these bases have been chosen, we can construct a direct-product basis
for ${\cal S}_{p}^{n}$: 
\begin{eqnarray*}
&&E_{i_{1}\ldots i_{n}\left( l_{1}\mu _{1}\right) \ldots \left( l_{p}\mu
_{p}\right) }^{\alpha _{1}\ldots \alpha _{n}}\left( {\bf \hat{r}}_{1},\ldots
,{\bf \hat{r}}_{p}\right) \\
&\equiv &e_{i_{1}}^{\alpha _{1}}\cdot \dots \cdot e_{i_{n}}^{\alpha
_{n}}\cdot Y_{l_{1},\mu _{1}}(\hat{{\bf r}}_{1})\cdot \dots \cdot
Y_{l_{p},\mu _{p}}(\hat{{\bf r}}_{p})
\end{eqnarray*}
The unitary matrix that connects the ${\bf E}_{i_{1}\ldots i_{n}\left(
l_{1}\mu _{1}\right) \ldots \left( l_{p}\mu _{p}\right) }$ basis to the $
{\bf B}_{qjm}$ basis can be calculated using the appropriate Clebsch-Gordan
coefficients. The calculation is straightforward but very long and tedious. 
However, the above analysis enables us to count
and classify the different irreducible representations of a given $j$. By
using the standard rules of angular-momentum addition: 
\[
s\otimes l=\left| s-l\right| \oplus \ldots \oplus \left( s+l\right) 
\]
we can count the number of irreducible representations of a given $j$. For
example, consider the space ${\cal S}_{1}^{2}$ of second-rank tensors with
one variable over the unit sphere. Using the angular-momentum addition rules
we see: 
\begin{eqnarray}
{\cal S}_{1}^{2} &=&1\otimes 1\otimes \left( 0\oplus 1\oplus 2\oplus 3\oplus
\ldots \right)  \label{dps21} \\
&=&\left( 0\oplus 1\oplus 2\right) \otimes \left( 0\oplus 1\oplus 2\oplus
3\oplus \ldots \right)  \nonumber \\
&=&\left( 3\times 0\right) \oplus \left( 7\times 1\right) \oplus \left(
9\times 2\right) \oplus \left( 9\times 3\right) \oplus \ldots  \nonumber
\end{eqnarray}
We see that there are exactly three $j=0$ representations, seven $j=1$
representations and 9 representations for each $j>1$. It can be further
argued that the symmetry properties of the basis tensors with respect to
their indices come from the $1\otimes 1=0\oplus 1\oplus 2$ part of the
direct product (\ref{dps21}). Therefore, out of the 9 irreducible
representation of $j>1$, 5 will be symmetric and traceless, 3 will be
anti-symmetric and 1 will be tracefull and diagonal. Similarly, the parity
of the resulting tensors (with respect to the single variable) can be
calculated.

Once we know how many irreducible representations of each $j$ are found
in ${\cal S}_{p}^{n}$, we can construct them ``by hand'', in some other,
more practical method which will be demonstrated in the following
subsection. 
\paragraph{Alternative derivation of the ${\bf B}_{qjm}$}.

\label{alternative}

The method we wish to propose in this subsection is based on the simple idea
that contractions with $r^{\alpha },\delta ^{\alpha \beta },\epsilon
^{\alpha \beta \gamma }$ and differentiation with respect to $r^{\alpha }$
are all {\em isotropic} operations. Isotropic in the sense that the
resulting expression will have the {\em same} transformation properties
under rotation as the expression we started with. The proof of the last
statement follows directly from the transformation properties of $r^{\alpha
},\delta ^{\alpha \beta },\epsilon ^{\alpha \beta \gamma }$.

The construction of all ${\bf B}_{qjm}$ that belongs to ${\cal S}_{1}^{n}$
now becomes a rather trivial task. We begin by defining a scalar tensor
field with a definite $j,m$. An obvious choice will be the well-known
spherical harmonics $Y_{jm}(\hat{{\bf r}})$, but a better one will be: 
\[
\Phi _{jm}({\bf r})\equiv r^{j}Y_{jm}(\hat{{\bf r}}). 
\]
The reason that we prefer $\Phi _{jm}({\bf r})$ to $Y_{jm}(\hat{{\bf r}})$,
is that $\Phi _{jm}({\bf r})$ is polynomial in ${\bf r}$ (while $Y_{jm}(\hat{
{\bf r}})$ is polynomial in $\hat{{\bf r}}$) and therefore it is easier to
differentiate it with respect to ${\bf r}$. Once we have defined $\Phi _{jm}(
{\bf r})$, we can construct the ${\bf B}_{qjm}$ by ``adding indices'' to $
\Phi _{jm}({\bf r})$ using the isotropic operations mentioned above. For
example, we may now construct:

\begin{itemize}
\item  $r^{-j}\delta ^{\alpha \beta }\Phi _{jm}({\bf r})$,

\item  $r^{-j+2}\delta ^{\alpha \beta }\partial ^{\tau }\partial ^{\gamma
}\Phi _{jm}({\bf r})$,

\item  $r^{-j-1}x^{\alpha }\Phi _{jm}({\bf r})$, \ etc...
\end{itemize}

Notice that we should always multiply the resulting expression with an
appropriate power of $r$, in order to make it $r$-independent, and thus a
function of $\hat{{\bf r}}$ only.

The crucial role of the Clebsch-Gordan analysis is to tell us how many
representations from each type we should come up with. First, it tells us
the highest power of $\hat{{\bf r}}$ in each representation, and then it can
also give us the symmetry properties of ${\bf B}_{qjm}$ with respect to their
indices. For example, consider the irreducible representations of ${\cal S}
_{1}^{2}$ - second rank tensors which depend on one unit vector $\hat{{\bf r}
}$. The Clebsch-Gordan analysis shows us that this space contains the
following irreducible representations: 
\begin{eqnarray*}
{\cal S}_{1}^{2} &=&{\cal S}_{0}^{1}\otimes {\cal S}_{0}^{1}\otimes {\cal S}
_{1}^{0} \\
&=&1\otimes 1\otimes \left( 0\oplus 1\oplus 2\oplus 3\oplus \ldots \right) \\
&=&\left( 0\oplus 1\oplus 2\right) \otimes \left( 0\oplus 1\oplus 2\oplus
3\oplus \ldots \right) \\
&=&\left( 3\times 0\right) \oplus \left( 7\times 1\right) \oplus \left(
9\times 2\right) \oplus \left( 9\times 3\right) \oplus \ldots
\end{eqnarray*}
That is, for each $j>1$ we're going to have 9 irreducible representations.
The indices symmetry of the tensor comes from the ${\cal S}_{0}^{1}\otimes 
{\cal S}_{0}^{1}=1\otimes 1=0\oplus 1\oplus 2$ part of the direct product.
This is a direct product of two Euclidean spaces, so its a second rank constant
tensor. We can mark the representation number in this space with the letter $
s$, and the representation number of the ${\cal S}_{1}^{0}=0\oplus 1\oplus
2\oplus 3\oplus \ldots $ space with the letter $l$. This way each
representation in ${\cal S}_{1}^{2}$ of a given $j$ will have two additional
numbers $(s,l)$, which actually serve as the index $q$ that distinguishes
irreducible representations of the same $j$. The $s$ index will determine
the indices symmetry of the tensor, while the $l$ index will determine the
highest power of $\hat{{\bf r}}$ in the tensor. If we now recall that in the
space of constant second-rank tensors, ${\cal S}_{0}^{1}\otimes {\cal S}
_{0}^{1}=0\oplus 1\oplus 2$, the $s=0,2$ representations are symmetric while
the $s=1$ representation is anti-symmetric, we can easily construct the $
{\bf B}_{qjm}$: 
\begin{equation}
\begin{array}{ll}
\left( s,l\right) =\left( 0,j\right) & B_{1jm}(\hat{{\bf r}})\equiv
r^{-j}\delta ^{\alpha \beta }\Phi _{jm}({\bf r}), \\ 
\left( s,l\right) =\left( 1,j-1\right) & B_{2jm}(\hat{{\bf r}})\equiv
r^{-j+1}\epsilon ^{\alpha \beta \mu }\partial _{\mu }\Phi _{jm}({\bf r}), \\ 
\left( s,l\right) =\left( 1,j\right) & B_{3jm}(\hat{{\bf r}})\equiv r^{-j}
\left[ r^{\alpha }\partial ^{\beta }-r^{\beta }\partial ^{\alpha }\right]
\Phi _{jm}({\bf r}), \\ 
\left( s,l\right) =\left( 1,j+1\right) & B_{4jm}(\hat{{\bf r}})\equiv
r^{-j-1}\epsilon ^{\alpha \beta \mu }r_{\mu }\Phi _{jm}({\bf r}), \\ 
\left( s,l\right) =\left( 2,j-2\right) & B_{5jm}(\hat{{\bf r}})\equiv
r^{-j+2}\partial ^{\alpha }\partial ^{\beta }\Phi _{jm}({\bf r}), \\ 
\left( s,l\right) =\left( 2,j-1\right) & B_{6jm}(\hat{{\bf r}})\equiv
r^{-j+1}\left[ \epsilon ^{\alpha \mu \nu }r_{\mu }\partial _{\nu }\partial
^{\beta }+\epsilon ^{\beta \mu \nu }r_{\mu }\partial _{\nu }\partial
^{\alpha }\right] \Phi _{jm}({\bf r}), \\ 
\left( s,l\right) =\left( 2,j\right) & B_{7jm}(\hat{{\bf r}})\equiv r^{-j}
\left[ r^{\alpha }\partial ^{\beta }+r^{\beta }\partial ^{\alpha }\right]
\Phi _{jm}({\bf r}), \\ 
\left( s,l\right) =\left( 2,j+1\right) & B_{8jm}(\hat{{\bf r}})\equiv
r^{-j-1}\left[ r^{\beta }\epsilon ^{\alpha \mu \nu }r_{\mu }\partial _{\nu
}+r^{\alpha }\epsilon ^{\beta \mu \nu }r_{\mu }\partial _{\nu }\right] \Phi
_{jm}({\bf r}), \\ 
\left( s,l\right) =\left( 2,j+2\right) & B_{9jm}(\hat{{\bf r}})\equiv
r^{-j-2}r^{\alpha }r^{\beta }\Phi _{jm}({\bf r}).
\end{array}
\label{eq:second-rank-tensors}
\end{equation}
It should be stressed that these ${\bf B}_{qjm}$ are not exactly the same
one we would have gotten from the Clebsch-Gordan machinery. For example,
they are not orthogonal among themselves for the same values of $j,m$
(although, they are orthogonal for different values of $j$ or $m$).
Nevertheless, they are linearly independent and thus span a given $(j,m)$
sector in the ${\cal S}_{1}^{2}$ space.

\section{The isotropy of the hierarchy of equations and its consequences}

In this section we derive the equations of motion for the statistical
averages of the velocity and pressure fields differences. We start from the
Navier-Stokes equations, and show that its isotropy implies the isotropy of
the equations for the statistical objects. Finally, we demonstrate the
foliation of these equations to different sectors of $j,m$.

Consider a Navier-Stokes incompressible turbulence in a bounded domain $
\Omega $. The equations of motion describing the flow are: 
\begin{eqnarray*}
\partial _{t}u^{\alpha }+u^{\mu }\partial _{\mu }u^{\alpha } &=&-\partial
^{\alpha }p+\nu \partial ^{2}u^{\alpha } \ , \\
\partial _{\alpha }u^{\alpha } &=&0\ .
\end{eqnarray*}
As is well known, the relevant dynamical time scales are revealed only when
the effect of sweeping is removed. In our work we use the Belinicher-L'vov
transformation \cite{87BL} in which the flow is observed from the point of
view of one specific fluid particle which is located at ${\bf r}_{0}$ at
time $t_{0}$. Let ${\bf \rho }({\bf r}_{0},t_{0}|t)$ be the particle's
translation at time $t$: 
\begin{equation}
{\bf \rho }({\bf r}_{0},t_{0}|t)=\int\limits_{t_{0}}^{t}ds{\bf u}[{\bf r}
_{0}+{\bf \rho }({\bf r}_{0},t_{0}|s),s]\ .  \label{a3}
\end{equation}
We then redefine the velocity and pressure fields to be those seen from an 
inertial frame whose origin sits at the current particle's position: 
\begin{eqnarray*}
{\bf v}({\bf r}_{0},t_{0}|{\bf r},t) &\equiv &{\bf u}[{\bf r}+{\bf \rho }(
{\bf r}_{0},t_{0}|t),t]\ , \\
\pi ({\bf r}_{0},t_{0}|{\bf r},t) &\equiv &p[{\bf r}+{\bf \rho }({\bf r}
_{0},t_{0}|t),t]\ .
\end{eqnarray*}
Next, we define the differences of these fields: 
\begin{eqnarray*}
{\cal W}^{\alpha }({\bf r}_{0},t_{0}|{\bf r},{\bf r}^{\prime },t) &\equiv
&v^{\alpha }({\bf r}_{0},t_{0}|{\bf r},t)-v^{\alpha }({\bf r}_{0},t_{0}|{\bf 
r}^{\prime },t)\ , \\
\Pi ({\bf r}_{0},t_{0}|{\bf r},{\bf r}^{\prime },t) &\equiv &\pi (
{\bf r}_{0},t_{0}|{\bf r},t)-\pi ({\bf r}_{0},t_{0}|{\bf r}^{\prime },t)\ .
\end{eqnarray*}
A straightforward calculation shows that the dynamical equations for $
{\bf {\cal W}}$ are: 
\begin{eqnarray}
\partial _{t}{\cal W}^{\alpha }({\bf r},{\bf r}^{\prime },t) &=&-\left(
\partial ^{\alpha }+\partial ^{\prime \alpha }\right) \Pi ({\bf r}
_{0},t_{0}|{\bf r},{\bf r}^{\prime },t)  \label{eq:WDyn} \\
&&+\nu \left( \partial ^{2}+\partial ^{\prime 2}\right) {\cal W}^{\alpha }(
{\bf r}_{0},t_{0}|{\bf r,r}^{\prime },t)  \nonumber \\
&&-\partial _{\mu }{\cal W}^{\mu }({\bf r}_{0},t_{0}|{\bf r,r}_{0},t){\cal W}
^{\alpha }({\bf r}_{0},t_{0}|{\bf r,r}^{\prime },t)  \nonumber \\
&&-\partial _{\mu }^{\prime }{\cal W}^{\mu }({\bf r}_{0},t_{0}|{\bf r}
^{\prime }{\bf ,r}_{0},t){\cal W}^{\alpha }({\bf r}_{0},t_{0}|{\bf r,r}
^{\prime },t)\ ,  \nonumber \\
\partial _{\alpha }{\cal W}^{\alpha }({\bf r}_{0},t_{0}|{\bf r},{\bf r}
^{\prime },t) &=&\partial _{\alpha }^{\prime }{\cal W}^{\alpha }({\bf r}
_{0},t_{0}|{\bf r},{\bf r}^{\prime },t)=0\ .  \nonumber
\end{eqnarray}
By inspection, $t_{0}$ merely serves as a parameter, and therefore we will
not denote it explicitly in the following discussion. Also, in order to make
the equations easier to understand, let us introduce some shorthand notation
for the variables $({\bf r}_{k},{\bf r}_{k}^{\prime },t_{k})$: 
\begin{eqnarray*}
{\bf X}_{k} &\equiv &({\bf r}_{k},{\bf r}_{k}^{\prime },t_{k})\ , \\
X_{k} &\equiv &(r_{k},r_{k}^{\prime },t_{k})\ , \\
{\hat{{\bf X}}}_{k} &\equiv &(\hat{{\bf r}}_{k},\hat{{\bf r}}_{k}^{\prime
})\ .
\end{eqnarray*}

Using (\ref{eq:WDyn}), we can now derive the dynamical equations for the
statistical moments of ${\bf {\cal W}}$,$\Pi $: Let $\left\langle \cdot
\right\rangle $ denote a suitable ensemble averaging. We define two types of
statistical moments: 
\begin{eqnarray*}
&&{\cal F}_{n}^{\alpha _{1}\ldots \alpha _{n}}({\bf r}_{0}|{\bf X}_{1},\ldots ,{\bf 
X}_{n}) \\
&\equiv &\left\langle {\cal W}^{\alpha _{1}}({\bf r}_{0}|{\bf X}_{1})\ldots 
{\cal W}^{\alpha _{n}}({\bf r}_{0}|{\bf X}_{n})\right\rangle \ , \\
&&{\cal H}_{n}^{\alpha _{2}\ldots \alpha _{n}}({\bf r}_{0}|{\bf X}_{1},\ldots ,{\bf 
X}_{n}) \\
&\equiv &\left\langle \Pi (r_{0}|{\bf X}_{1}){\cal W}^{\alpha _{2}}({\bf r}
_{0}|{\bf X}_{2})\ldots {\cal W}^{\alpha _{n}}({\bf r}_{0}|{\bf X}
_{n})\right\rangle \ .
\end{eqnarray*}
Equation (\ref{eq:WDyn}) implies: 
\begin{eqnarray}
&&\partial _{t_{1}}{\cal F}_{n}^{\alpha _{1}\ldots \alpha _{n}}({\bf r}_{0}|{\bf X}
_{1},\ldots ,{\bf X}_{n})  \label{eq:FDyn} \\
&=&-\left( \partial _{(r_{1})}^{\alpha _{1}}+\partial _{(r_{1}^{\prime
})}^{\alpha _{1}}\right) {\cal H}_{n}^{\alpha _{2}\ldots \alpha _{n}}({\bf r}_{0}|
{\bf X}_{1},\ldots ,{\bf X}_{n})  \nonumber \\
&&-\partial _{\mu }^{(r_{1})}{\cal F}_{n+1}^{\mu \alpha _{1}\ldots \alpha
_{n}}\left( {\bf r}_{0}|{\tilde{{\bf X}},X}_{1},\ldots ,{\bf X}_{n}\right)  
\nonumber \\
&&-\partial _{\mu }^{(r_{1}^{\prime })}{\cal F}_{n+1}^{\mu \alpha _{1}\ldots \alpha
_{n}}\left( {\bf r}_{0}|{\tilde{{\bf X}}}^{\prime }{\bf ,X}_{1},\ldots ,{\bf 
X}_{n}\right)   \nonumber \\
&&+\nu \left( \partial _{(r_{1})}^{2}+\partial _{(r_{1}^{\prime
})}^{2}\right) {\cal F}_{n}^{\alpha _{1}\ldots \alpha _{n}}({\bf r}_{0}|{\bf X}
_{1},\ldots ,{\bf X}_{n})\ ,  \nonumber \\
{\tilde{{\bf X}}} &\equiv &{(}{\bf r}_{0},{\bf r}^{\prime },t)\;;\;{\tilde{
{\bf X}}}^{\prime }{\equiv (}{\bf r},{\bf r}_{0},t)\ ,
\end{eqnarray}
\begin{eqnarray}
\partial _{\alpha _{1}}^{(r_{1})}{\cal F}_{n}^{\alpha _{1}\ldots \alpha _{n}}({\bf r
}_{0}|{\bf X}_{1},\ldots ,{\bf X}_{n}) &=&0\ ,  \label{eq:FIncomp} \\
\partial _{\alpha _{1}}^{(r_{1}^{\prime })}{\cal F}_{n}^{\alpha _{1}\ldots \alpha
_{n}}({\bf r}_{0}|{\bf X}_{1},\ldots ,{\bf X}_{n}) &=&0\ .  \nonumber
\end{eqnarray}

Equations (\ref{eq:FDyn}), (\ref{eq:FIncomp}) are linear and homogeneous.
Therefore their solutions form a linear space. The most general solution to
these equations is given by a linear combination of a suitable basis of the
solutions space. To construct a specific solution, we must use the boundary
conditions in order to set the linear weights of the basis solutions. We
shall now show that the isotropy of these equations implies that our
basis of solutions can be constructed such that every solution will have a
definite behavior under rotations (that is, definite $j$ and $m$ - see Sect.
2). But before we do that, note that in many aspects the situation
described here is similar to the well-known problem of Laplace equation in a
closed domain $\Omega $: 
\begin{eqnarray*}
\nabla ^{2}\Psi &=&0\ , \\
\Psi |_{\partial \Omega } &=&\sigma \ .
\end{eqnarray*}
The Laplace equation is linear, homogeneous and isotropic. Therefore its
solutions form a linear space. One possible basis for this space is: 
\[
\Psi _{l,m}({\bf r})\equiv r^{l}Y_{lm}(\hat{{\bf r}})\ , 
\]
in which the solutions have a definite behavior under rotations (belong
to an irreducible representation of $SO(3)$ ). The general solution of the
problem is given as a linear combination of the $\Psi _{l,m}({\bf r})$: 
\[
\Psi ({\bf r})=\sum_{l,m}a_{l,m}\Psi _{l,m}({\bf r})\ . 
\]
For a specific problem, we use the value of $\Psi ({\bf r})$ on the boundary
(i.e., we use $\sigma ({\bf r})$) in order to set the values of $a_{l,m}$.

To see that the same thing happens with the hierarchy equations (\ref
{eq:FDyn}, \ref{eq:FIncomp}), we consider an arbitrary solution $\{{\B.{\cal F}}
_{n},{\B.{\cal H}}_{n}|\;n=2,3,\ldots \}$ of these equations. According to Sect.2
we may write the tensor fields ${\B.{\cal F}}_{n},{\B.{\cal H}}_{n}$ in terms of a
basis ${\bf B}_{qjm}$: 
\begin{eqnarray}
&&{\cal F}_{n}^{\alpha _{1}\ldots \alpha _{n}}({\bf r}_{0}|{\bf X}_{1},\ldots ,{\bf 
X}_{n})  \label{def:jmExpansion} \\
&\equiv &\sum_{q,j,m} F_{qjm}^{(n)}(r_{0},X_{1},\ldots ,X_{n})  \nonumber \\
&&\times B_{qjm}^{(n)\;\alpha _{1}\ldots \alpha _{n}}(\hat{{\bf r}}_{0},{
\hat{{\bf X}}}_{1},\ldots {\hat{{\bf X}}}_{n})\ ,  \nonumber
\end{eqnarray}
\begin{eqnarray*}
&&{\cal H}_{n}^{\alpha _{2}\ldots \alpha _{n}}({\bf r}_{0}|{\bf X}_{1},\ldots ,{\bf 
X}_{n}) \\
&\equiv &\sum_{q,j,m} H_{qjm}^{(n)}(r_{0},X_{1},\ldots ,X_{n}) \\
&&\times B_{qjm}^{(n-1)\;\alpha _{2}\ldots \alpha _{n}}(\hat{{\bf r}}_{0},{
\hat{{\bf X}}}_{1},\ldots {\hat{{\bf X}}}_{n})\ .
\end{eqnarray*}
Now all we have to show is that the pieces of ${\B.{\cal F}}_{n},{\B.{\cal H}}_{n}$ with
definite $j,m$ solve the hierarchy equations {\em by themselves} -
independently of pieces with different $j,m$. The proof of the last
statement is straightforward though somewhat tedious. We therefore only
sketch it in general lines. The isotropy of the hierarchy equations implies
that pieces of ${\B.{\cal F}}_{n},{\B.{\cal H}}_{n}$ with definite $j,m$, maintain their
transformations properties under rotation {\em after} the linear and
isotropic operations of the equation have been performed. For example, if $
{\cal F}_{n}^{\alpha _{1}\ldots \alpha _{n}}({\bf r}_{0}|{\bf X}_{1},\ldots ,{\bf X}
_{n})$ belongs to the irreducible representation $(j,m)$, then so will the
tensor fields: 
\[
\partial _{\alpha _{i}}^{(r_{k})}{\cal F}_{n}^{\alpha _{1}\ldots \alpha
_{n}},\;\partial _{(r_{k})}^{2}{\cal F}_{n}^{\alpha _{1}\ldots \alpha _{n}},\;\text{
etc...}
\]
although, they may belong to different ${\cal S}_{p}^{n}$ spaces (i.e., have
one less or one more indices). Therefore, if we choose the bases $\left\{ 
{\bf B}_{qjm}^{(n)}\right\} $ to be orthonormal, plug the expansion (\ref
{def:jmExpansion}) into the hierarchy equations equations (\ref{eq:FDyn}, 
\ref{eq:FIncomp}), and take the inner product with ${\bf B}_{qjm}^{(n)}$, we
will obtain new equations for the scalar functions $F_{qjm}^{(n)},H_{qjm}^{(n)}$:
\begin{eqnarray}
&&\partial _{t_{1}}F_{qjm}^{(n)}(r_{0},X_{1},\ldots ,X_{n})
\label{eq:ScalarDyn} \\
&=&-\sum_{q^{\prime }}\left\langle \left( \partial _{(r_{1})}^{\alpha
_{1}}+\partial _{(r_{1}^{\prime })}^{\alpha _{1}}\right) H_{q^{\prime
}jm}^{(n)}(r_{0},X_{1},\ldots ,X_{n}){\bf B}_{q^{\prime }jm}^{(n-1)},{\bf B}
_{qjm}^{(n)}\right\rangle   \nonumber \\
&&-\sum_{q^{\prime }}\left\langle \partial _{\mu }^{(r_{1})}F_{q^{\prime
}jm}^{(n+1)}(r_{0},\tilde{X},X_{1},\ldots ,X_{n}){\bf B}_{q^{\prime
}jm}^{(n+1)},{\bf B}_{qjm}^{(n)}\right\rangle   \nonumber \\
&&-\sum_{q^{\prime }}\left\langle \partial _{\mu }^{(r_{1}^{\prime
})}F_{q^{\prime }jm}^{(n+1)}(r_{0},\tilde{X}^{\prime },X_{1},\ldots ,X_{n})
{\bf B}_{q^{\prime }jm}^{(n+1)},{\bf B}_{qjm}^{(n)}\right\rangle   \nonumber
\\
&&+\nu \sum_{q^{\prime }}\left\langle \left( \partial
_{(r_{1})}^{2}+\partial _{(r_{1}^{\prime })}^{2}\right) F_{q^{\prime
}jm}^{(n)}(r_{0},X_{1},\ldots ,X_{n}){\bf B}_{q^{\prime }jm}^{(n)},{\bf B}
_{qjm}^{(n)}\right\rangle ,  \nonumber
\end{eqnarray}
\begin{eqnarray}
\sum_{q^{\prime }}\left\langle \partial _{\alpha _{1}}^{(r_{1})}F_{q^{\prime
}jm}^{(n)}(r_{0},X_{1},\ldots ,X_{n}){\bf B}_{q^{\prime }jm}^{(n)},{\bf B}
_{qjm}^{(n-1)}\right\rangle  &=&0\ ,  \label{eq:ScalarIncomp} \\
\sum_{q^{\prime }}\left\langle \partial _{\alpha _{1}}^{(r_{1}^{\prime
})}F_{q^{\prime }jm}^{(n)}(r_{0},X_{1},\ldots ,X_{n}){\bf B}_{q^{\prime
}jm}^{(n)},{\bf B}_{qjm}^{(n-1)}\right\rangle  &=&0\ .  \nonumber
\end{eqnarray}
Note that in the above equations, $\left\langle \cdot \right\rangle $ denote
the inner-product in the ${\cal S}_{p}^{n}$ spaces. Also, the sums over $
q^{\prime },j^{\prime },m^{\prime }$ from (\ref{def:jmExpansion}) was
reduced to a sum over $q^{\prime }$ only - due to the isotropy. We thus see
explicitly from (\ref{eq:ScalarDyn},\ref{eq:ScalarIncomp}) the decoupling of
the equations for different $j,m$.

\subsection{Rescaling Symmetry and Anomalous Exponents}

The hierarchical equations simplify somewhat in the limit of infinite Reynolds
number Re$\to\infty$. This limit is equivalent to $\nu\to 0$, in which the last term
in Eqs.(\ref{eq:ScalarDyn}) can be neglected with impunity. It was pointed
out before \cite{99BLPP} that this is the main advantage of using ``fully unfused" 
correlation functions in which all the coordinates are distinct: there is
nothing to compensate for the vanishing of the viscosity in the $\nu\to 0$ limit.
Once the viscous term is discarded, the rest of the equations .
exhibit invariance to rescaling under the following rescaling group: 
\begin{eqnarray}
{\bf r}_i\to \lambda {\bf r}_i,\quad t_i\to \lambda^{1-h}t_1, \quad
&&F_{qjm}^{(n)}\to \lambda ^{nh+{\cal Z}_j(h)}F_{qjm}^{(n)} \ ,  \nonumber\\
&&H_{qjm}^{(n)}\to \lambda ^{(n+1)h+{\cal Z}_j(h)}H_{qjm}^{(n)} \ , \label{resa}
\end{eqnarray}
as can be verified by direct substitution. In (\ref{resa}) $\lambda$ and $h$
are arbitrary scalars, and ${\cal Z}_j(h)$ is an arbitrary $n$-independent
scalar function. We endow it with an index $j$ since we expect, and see below,
that  ${\cal Z}_j(h)$ will differ in different $j$-sectors, but not in
different $m$-sectors. 

As a consequence of the rescaling symmetry we can seek solutions that do not
mix values of $h$. We define $\tilde F_{qjm,h}^{(n)}$ and $\tilde H_{qjm,h}^{(n)}$
as the quantities that
solve the equations of motion on an $h$-slice, which are the same as equations
(\ref{eq:ScalarDyn}) without the viscous term. 
The important property of the solution on an $h$-slice is that it is a
homogeneous function of all its arguments in the sense that 
\begin{eqnarray}
&&\tilde F_{qjm,h}(\lambda r_0,\lambda r_1,\lambda r'_1,
\lambda^{1-h}t_1, \dots \lambda r_n, \lambda r'_n,
\lambda^{1-h}t_n)=  \nonumber \\
&&\lambda^{nh+{\cal Z}_j(h)} \tilde F_{qjm,h}(r_0, r_1,r'_1,t_1,\dots 
r_n, r'_n,t_n) \ .  \label{resah}
\end{eqnarray}
It should be stressed that the quantity $F_{qjm}^{(n)}$ itself is {\em not}
homogeneous in its arguments. It has been discovered in \cite{97LPP-2} and stressed
in \cite{99BLPP} that time-correlation functions in turbulence do not satisfy
dynamic scaling in the sense of Eq.(\ref{resah}). Indeed, the solution of
Eq.(\ref{eq:ScalarDyn}) is a sum over contributions on $h$-slices, 
\begin{equation}
F_{qjm}^{(n)}(r_0,X_1\dots ,X_n)=\int_{h_{{\rm min}}}^{h_{{\rm max}}}
d\mu(h) \tilde F_{qjm,h}^{(n)}(r_0,X_1,\dots ,X_n) \ ,  \label{sumh}
\end{equation}
with $\mu(h)$ some unknown measure that needs to be obtained from boundary
conditions. Eq.(\ref{sumh}) can be endowed with further meaning by rescaling
coordinates and times according to 
\begin{equation}
{\bf \rho}_j\equiv {\bf r}_j/R_n,~{\bf \rho}^{\prime}_j\equiv {\bf r}
^{\prime}_j/R_n,~\tau_j\equiv t_j/t_{R_n,h} \ ,  \label{scaledvar}
\end{equation}
where $R_n$ and $t_{R_n,h}$ are defined as the typical scale of separation
of the set of coordinates and the typical times scale on that scale on an $h$
-slice: 
\begin{eqnarray}
R^2_n&\equiv&{\frac{1}{n}}\sum_{j=1}^n |{\bf r}_j-{\bf r}^{\prime}_j|^2 \ ,
\label{Rn} \\
t_{R_n,h}&\equiv& {\frac{R}{U}}\left({\frac{L}{R}}\right)^h \ .  \label{tRh}
\end{eqnarray}
Here $U$ is the typical velocity on the outer scale of turbulence $L$.
Defining now 
\begin{equation}
\Xi_j\equiv ({\bf \rho}_j,{\bf \rho}^{\prime}_j,\tau_j) \ ,  \label{xi}
\end{equation}
Eq.(\ref{sumh}) can be written, using the rescaling property on an $h$ slice
as 
\begin{eqnarray}
&&F_{qjm}^{(n)}(X_1,\dots ,X_n) =  \nonumber \\
&&U^n\int_{h_{{\rm min}}}^{h_{{\rm max}}} d\mu(h)\left({\frac{R_n}{L}}
\right)^{nh+{\cal Z}_j(h)} \tilde F_{qjm,h}^{(n)}(\Xi_1,\dots ,\Xi_n)
\end{eqnarray}
This form is known as the ``multi-fractal" form \cite{85PF,86HJKPS,84BPPV}. The scaling exponents
characterizing $a_{qjm}^{(n)}$ are obtained from a saddle-point calculation
in the limit $R/L\to 0$ as ${\rm min}_h\{nh+{\cal Z}_j(h)\}$.

It was explained in \cite{99BLPP} that ${\cal Z}_j(h)$ is obtained from a
solvability condition of the hierarchy of equations (\ref{eq:ScalarDyn}). In
particular the numerical value of the function ${\cal Z}_j(h)$, and consequently
of the scaling exponents which are determined by the saddle point integral,
depend on the {\em coefficients} in the equations (\ref{eq:ScalarDyn}). We found
that the scalar functions associated with the different $j$ -irreducible
representations, $F_{qjm}^{(n)}(X_{1},\dots ,X _{n})$,
satisfy equations with different coefficients, depending on inner products of the
basis functions ${\bf B}_{qjm}$. Accordingly we expect the scalar function $
{\cal Z}_j(h)$ to change from sector to sector. If the functions $
F_{qjm}^{(n)} $ are characterized by anomalous exponents, they may be
different for different $j$. On the other hand, for the same $j$ the
equations mix different $m$ (and $q$) components, and unless there is an
additional symmetry to SO(3), we do not expect different contributions with
the same $j$ to exhibit different exponents. In the next section we will
demonstrate explicitly in the context of 3rd order correlation functions how
the existence of an additional symmetry, in that case parity, brings about a
foliation of a $j$ sector into two sub-sectors which exhibit two different
scaling exponents.
\section{Example: Kolmogorov's ``Four-Fifth law" and the foliation to
different $j$'s}

One of the best known results in the statistical theory of turbulence is
Kolmogorov's ``four-fifth law'' which was discovered in 1941 \cite{41Kol}.
This law pertains to the third order moment of longitudinal velocity
differences $\delta u_{l}({\bf r},{\bf R},t)\equiv \lbrack {\bf u}({\bf r}+ 
{\bf R},t)-{\bf u}({\bf r},t)]\cdot {\bf R}/R$ where ${\bf u}({\bf r},t)$ is
the Eulerian velocity field of the turbulent fluid. The fourth-fifth law
states that in homogeneous, isotropic and stationary turbulence, in the
limit of vanishing kinematic viscosity $\nu \rightarrow 0$ 
\begin{equation}
\left\langle \lbrack \delta u_{l}({\bf r},{\bf R},t)]^{3}\right\rangle =- 
\case{4}{5}\bar{\epsilon}R\ ,  \label{4/5}
\end{equation}
where $\bar{\epsilon}$ is the mean energy flux per unit time and mass $\bar{
\epsilon}\equiv \nu \left\langle |\nabla _{\alpha }u_{\beta
}|^{2}\right\rangle $. The only assumption needed to derive this law is that
the dissipation is finite in the limit $\nu \rightarrow 0$.

In this section we revisit this law by finding the full tensorial form of
the $j=0$ component of the 3rd order correlation function. We {\em do not
need} to assume isotropy of the turbulence at any stage of the development;
the isotropy of the equations of motion suffices to decouple the $j=0$
contribution from all the rest, and in this case we have enough equations to
determine the $j=0$ component of the tensor completely. We will also show
that the $j=0$ component has two subcomponents with different scaling
exponents. These subcomponents have different parity, and therefore are
further decoupled in the equations of motion. The usual fourth-fifth law
pertains to the components that have odd parity. One can derive an
additional exact relation that pertains to the even parity components \cite{CD-LPP}.

Defining the velocity ${\bf v}({\bf r},t)$ as ${\bf v}({\bf r},t)\equiv {\bf 
u}({\bf r},t)-\left\langle {\bf u}\right\rangle $ we consider the
simultaneous 3rd order tensor correlation function which depends on two
space points: 
\begin{equation}
J^{\alpha ,\beta \gamma }({\bf R})\equiv \left\langle v^{\alpha }({\bf r}+ 
{\bf R},t)v^{\beta }({\bf r},t)v^{\gamma }({\bf r},t)\right\rangle \ .
\label{defJ}
\end{equation}
We show \cite{CD-LPP} that in the limit $\nu \rightarrow 0$, under the same assumption
leading to the fourth-fifth law, this correlation function reads 
\begin{eqnarray}
J^{\alpha ,\beta \gamma }({\bf R})=- &&{\frac{\bar{\epsilon}}{10}}(R^{\gamma
}\delta ^{\alpha \beta }+R^{\beta }\delta ^{\alpha \gamma }-{\frac{2}{3}}
R^{\alpha }\delta ^{\beta \gamma })  \label{eq:s3j0} \\
&-&{\frac{\bar{h}}{30}}(\epsilon ^{\alpha \beta \delta }R^{\gamma }+\epsilon
^{\alpha \gamma \delta }R^{\beta })R_{\delta }\ .  \label{result}
\end{eqnarray}
The quantity $\bar{h}$ is the mean dissipation of helicity per unit mass and
time 
\begin{equation}
\bar{h}\equiv \nu \left\langle (\partial ^{\alpha }u^{\beta })(\partial
^{\alpha }[{\bf \nabla \times }{\bf u}]^{\beta })\right\rangle \ ,
\label{defh}
\end{equation}
The new result (\ref{result}) (derived firstly in \cite{CD-LPP} and \cite{98Rus}) can
be also displayed in a form that depends on $\bar{h}$ alone by introducing
the longitudinal and transverse parts of ${\bf u}$: the longitudinal part is 
${\bf u}_{l}\equiv {\bf R}({\bf u\cdot } {\bf R})/R^{2}$ and the transverse
part is ${\bf u}_{t}\equiv {\bf u}-{\bf u} _{l}$. In addition we have $
\delta {\bf u}_{l}({\bf r},{\bf R},t)\equiv \delta u_{l}({\bf r},{\bf R},t)
{\bf R}/R$. In terms of these quantities we can propose a ``two fifteenth
law'' that pertains to the $j=0$ component of the following correlation
function: 
\begin{equation}
\left\langle \lbrack \delta {\bf u}_{l}({\bf r},{\bf R},t)]\cdot \lbrack 
{\bf u}_{t}({\bf R}+{\bf r},t)\times {\bf u}_{t}({\bf r},t)]\right\rangle = 
\case{2}{15}\bar{h}R^{2}\ .  \label{short2:15}
\end{equation}
We note that this result holds also when we replace ${\bf u}$ by ${\bf v}$
everywhere.

To derive the result (\ref{result}) we start from the correlation function $
J^{\alpha ,\beta \gamma }({\bf R})$ which is symmetric with respect to
exchange of the indices $\beta $ and $\gamma $ as is clear from the
definition. Using the symmetry the most general form of the $j=0$ component
of this tensor can be written by observation (Sect. 2b): 
\begin{eqnarray}
&&J^{\alpha ,\beta \gamma }({\bf R})=a_{1}(R)[\delta ^{\alpha \beta }\hat{R}
^{\gamma }+\delta ^{\alpha \gamma }\hat{R}^{\beta }+\delta ^{\beta \gamma }
\hat{R}^{\alpha }]  \label{general} \\
&&+a_{2}(R)[\delta ^{\alpha \beta }\hat{R}^{\gamma }+\delta ^{\alpha \gamma }
\hat{R}^{\beta }-2\delta ^{\beta \gamma }\hat{R}^{\alpha }]  \nonumber \\
&&+a_{3}(R)[\delta ^{\alpha \beta }\hat{R}^{\gamma }+\delta ^{\alpha \gamma }
\hat{R}^{\beta }+\delta ^{\beta \gamma }\hat{R}^{\alpha }-5R^{\alpha }\hat{R}
^{\beta }\hat{R}^{\gamma }/R^{2}]  \nonumber \\
&&+a_{4}(R)[\epsilon ^{\alpha \beta \delta }\hat{R}^{\gamma }+\epsilon
^{\alpha \gamma \delta }\hat{R}^{\beta }]\hat{R}_{\delta }
\end{eqnarray}
This form is precisely of the type $\sum_{q}a_{qjm}{\bf B}_{qjm}$ for the
isotropic sector $j=m=0$.

Not all the coefficients are independent for incompressible flows. Requiring 
$\partial J^{\alpha,\beta\gamma}({\bf R})/\partial R^\alpha=0$ leads to two
relations among the coefficients: 
\begin{eqnarray}
&&\Big({\frac{d}{dR}}+{\frac{4}{R}}\Big)a_3(R)={\frac{2}{3}}[{\frac{d}{dR}}-{
\ \frac{1}{R}}]\big[a_1(R) + a_2(R)\big] \ ,  \label{increl} \\
&&\Big({\frac{d}{dR}}+{\frac{2}{R}}\Big)\big[5a_1(R)-4 a_2(R)\big]=0 \ . 
\nonumber
\end{eqnarray}
As we have two conditions relating the three coefficients $a_1,~a_2,~a_3$
only one of them is independent. Kolmogorov's derivation \cite{41Kol}
related the rate of energy dissipation to the value of the remaining
unknown. Here the coefficient $a_4$ remains undetermined by the
incompressibility constraint. It belongs to a component of odd parity;
since the equations of motion and the incompressibility constraint are
invariant under parity transformation it decouples altogether and needs to be
determined separately.

Kolmogorov's derivation can be paraphrased in a simple manner. Begin with
the second order structure function which is related to the energy of $R$
-scale motions 
\begin{equation}
S_{2}(R)\equiv \left\langle |{\bf u}({\bf R}+{\bf r},t)-{\bf u}({\bf r}
,t)|^{2}\right\rangle \ .  \label{S2}
\end{equation}
Computing the rate of change of this (time-independent) function from the
Navier-Stokes equations we find 
\begin{equation}
0={\frac{\partial S_{2}(R)}{2\partial t}}=-{\cal D}_{2}(R)-2\bar{\epsilon}
+\nu \nabla ^{2}S_{2}(R)\ ,  \label{bal}
\end{equation}
where ${\cal D}_{2}(R)$ stems from the nonlinear term $({\bf u}\cdot \nabla
) {\bf u}$ and as a result it consists of a correlation function including a
velocity derivative. The conservation of energy allows the derivative to be
taken outside the correlation function: 
\begin{eqnarray}
{\cal D}_{2}(R) &\equiv &{\frac{\partial }{\partial R^{\beta }}}\langle
u^{\alpha }({\bf r},t)u^{\alpha }({\bf r}+{\bf R},t)\big[u^{\beta }({\bf r}
,t)  \label{D2} \\
&&-\!u^{\beta }({\bf r}+{\bf R},t)\big]\rangle \ .  \nonumber
\end{eqnarray}
In terms of the function of Eq. (\ref{defJ}) we can write 
\begin{equation}
{\cal D}_{2}(R)={\frac{\partial }{\partial R^{\beta }}}\Big[J^{\alpha ,\beta
\alpha }({\bf R},t)-J^{\alpha ,\beta \alpha }(-{\bf R},t)\Big]\ .
\label{relate}
\end{equation}
Note that Eq. (\ref{defJ}) is written in terms of ${\bf v}$ rather than $
{\bf u}$, but using the incompressibility constraint we can easily prove
that Eq. (\ref{D2}) can also be identically written in terms of ${\bf v}$
rather than ${\bf u}$. We proceed using Eq. (\ref{general}) in Eq. (\ref
{relate}), and find 
\begin{equation}
{\cal D}_{2}(R)=2{\frac{\partial }{\partial R^{\beta }}}\hat{R}^{\beta }
\big[ 5a_{1}(R)+2\tilde{a}_{1}(R)\big]\ .  \label{D2a1}
\end{equation}
For $R$ in the inertial interval, and for $\nu \rightarrow 0$, we can read
from Eq. (\ref{bal}) ${\cal D}_{2}(R)=-2\bar{\epsilon}$ and therefore have
the third relation that is needed to solve all the three unknown
coefficients. A calculation leads to 
\begin{equation}
a_{1}(R)=-2\bar{\epsilon}R/45\ ,\quad a_{2}=-\bar{\epsilon}R/18\ ,\quad
a_{3}=0\ .  \label{avalues}
\end{equation}
The choice of the structure function $S_{2}(R)$ leads to Eq.(\ref{relate})
in which the odd parity components disappear, leaving $a_{4}(R)$
undetermined. Another correlation function is needed in order to remedy the
situation. Since the helicity is ${\bf u\cdot }[{\bf \nabla \times }{\bf u}]$
, we seek a correlation function which is related to the helicity of eddies
of scale of $R$: 
\begin{eqnarray}
&&T_{2}(R)\equiv \langle \big[{\bf u}({\bf R}+{\bf r},t)-{\bf u}({\bf r},t) 
\big]  \nonumber \\
&&\cdot \big[{\bf \nabla \times }{\bf u}({\bf r}+{\bf R},t)-{\bf \nabla
\times }{\bf u}({\bf r},t)\big]\rangle \ .  \label{T2}
\end{eqnarray}
The proper choice of this correlation function is the crucial idea here. The
rest is a straightforward calculation. Using the Navier-Stokes equations to
compute the rate of change of this quantity we find 
\begin{equation}
0={\frac{\partial T_{2}(R)}{2\partial t}}=-G_{2}(R)-2\bar{h}-\nu \nabla
^{2}T_{2}(R)\ ,  \label{bal2}
\end{equation}
which is the analog of (\ref{bal}), and where 
\begin{eqnarray}
&&G_{2}(R)=\{\langle {\bf u}({\bf r},t)\cdot \big[{\bf \nabla }_{R}\times 
\big[{\bf u}({\bf r}+{\bf R},t)\times \big[{\bf \nabla }_{R}  \label{G2} \\
&&\times {\bf u}({\bf r}+{\bf R},t)\big]\big]\big]\rangle \}+\{{\rm term}~ 
{\bf R\rightarrow }-{\bf R\}}\ .  \nonumber
\end{eqnarray}
The conservation of helicity allows the extraction of two derivatives
outside the correlation functions. The result can be expressed in terms of
our definition (\ref{defJ}): 
\begin{equation}
G_{2}(R)={\frac{\partial }{\partial R^{\lambda }}}{\frac{\partial }{\partial
R^{\kappa }}}\epsilon _{\alpha \lambda \mu }\epsilon _{\mu \beta \nu
}\epsilon _{\nu \kappa \gamma }\big[J^{\alpha ,\beta \gamma }({\bf R}
)+J^{\alpha ,\beta \gamma }(-{\bf R})\big]\ .  \label{G2J}
\end{equation}
Substituting Eq. (\ref{general}) we find 
\begin{equation}
G_{2}(R)=2{\frac{\partial ^{2}}{\partial R^{\lambda }\partial R^{\kappa }}}
b_{2}(R)\big[\hat{R}^{\lambda }\hat{R}^{\kappa }-\delta ^{\lambda \kappa } 
\big]\ ,  \label{G2b}
\end{equation}
which is the analog of Eq. (\ref{D2a1}). Using Eq. (\ref{bal2}) in the
inertial interval in the limit $\nu \rightarrow 0$ we find the differential
equation 
\begin{equation}
{\frac{d^{2}a_{4}(R)}{dR^{2}}}+5{\frac{1}{R}}{\frac{db_{2}(R)}{dR}}+{\frac{
3a_{4}(R)}{R^{2}}}=-{\frac{\bar{h}}{2}}\ .  \label{difb2}
\end{equation}
The general solution of this equation is 
\begin{equation}
a_{4}(R)=-\bar{h}R^{2}/30+\alpha _{1}R^{-3}+\alpha _{2}R^{-1}\ .
\end{equation}
Requiring finite solutions in the limit $R\rightarrow 0$ means that $\alpha
_{1}=\alpha _{2}=0$. Accordingly we end up with Eq. (\ref{result}). 
We restate again, that in the preceding derivation, we did not assume that
our turbulence were isotropic. Other terms of $J^{\alpha ,\beta \gamma }({\bf 
R} )$ with $j\neq 0$ can possibly contribute to the total $J^{\alpha ,\beta
\gamma }({\bf R})$. However, under the assumptions of homogeneity and finite
energy and helicity dissipations, the $R$ dependence of the
isotropic part of $J^{\alpha ,\beta \gamma }({\bf R})$ is necessarily as
stated in (\ref{result}).

It should be noted, that a parallel calculation can be easily carried out
for the $j>0$ sectors of the third order correlation function $J^{\alpha
,\beta \gamma }({\bf R})$. In these sectors however, there are more
irreducible representations than there are in the $j=0$ sector (to be exact,
there is a total of 18 representations for each $j>2$. 10 of them with $
(-)^{j+1}$ parity and 8 of them with $(-)^{j}$ parity). As a result, for $
j>0 $ we get more unknown functions $a_{qjm}(R)$ than equations, and hence
we cannot obtain a full solution. Our failure in obtaining a complete set of
equations for $j>0$ sectors, can be attributed to the inadequacy of a
turbulence theory that involves only moments which are simultaneous in time. 
The $j=0$ sector is indeed unique in the sense that its low number of
irreducible representations makes a full solution possible. 

\section{Example: analysis of anisotropy in atmospheric turbulence}

\label{sec:experiment}

In this section we present experimental evidence for the utility of the
expansion (\ref{Texp}) in terms of the irreducible representations. The
analysis of the experimental data was done in collaboration with B. Dhruva,
S. Kurien and K.R. Sreenivasan, and the reader is referred to the details of 
\cite{99ADKLPS}. In that work we focused on the 2nd rank tensor structure functions
of velocity differences 
\begin{equation}
S^{\alpha \beta }({\bf R})\equiv \left\langle \left[ u^{\alpha }({\bf r}+
{\bf R})-u^{\alpha }({\bf r})\right] \left[ u^{\beta }({\bf r}+{\bf R}
)-u^{\beta }({\bf r})\right] \right\rangle \ .  \label{sf}
\end{equation}
where {\em homogeneity} of the flow is assumed, but not isotropy. This
object is symmetric in its indices and has even parity in ${\bf R}$. In
addition it is expected to scale with ${\bf R}$ in the inertial range, with
possibly different scaling exponent characterizing contributions of
different $j$. We demonstrated in \cite{99ADKLPS} that one can usefully represent $
S^{\alpha \beta }({\bf R})$ in the form: 
\begin{equation}
S^{\alpha \beta }({\bf R})=\sum_{qjm}a_{qjm}|{\bf R}|^{\zeta
_{2}^{(j)}}B_{qjm}^{\alpha \beta }({\bf \hat{R}})  \label{trial}
\end{equation}
where $a_{qjm}$ are some numerical coefficients, $B_{qjm}^{\alpha \beta }(
{\bf \hat{R}})$ are the tensor basis of ${\cal S}_{1}^{2}$ with a definite $
j,m$, and $\zeta _{2}^{(j)}$ are the exponents associated with the $j$'s
irreducible representation. The isotropic exponent, $\zeta _{2}^{(0)}$, will
be referred to shortly as $\zeta _{2}$. We note that the coefficients $a_{qjm}$
are {\em not} arbitrary numerical coefficients, because of the constraints
imposed by the incompressibility of the flow. In Appendix \ref
{sec:general-form} we derive the explicit form of $B_{qjm}^{\alpha \beta }(
{\bf \hat{R}})$ and the necessary relations among the $a_{qjm}$'s. The
theoretical development of Appendix \ref{sec:general-form} serves as a basis
for the data analysis; we leave it in the appendix since it is somewhat
lengthy. Nevertheless the interested reader may find it useful for
situations that differ from the one treated below.

The data that we want to consider were taken at Taylor microscale Reynolds
numbers of about 10,000 \cite{99ADKLPS}. The data were acquired simultaneously from two
single-wire probes separated by $\Delta =55$ cm nominally orthogonal to the
mean wind direction. The two probes were mounted at a height of 6 m
over a flat desert with a long fetch. The Kolmogorov scale was about 0.75 cm. Details of the
experimental setup can be found in ref. \cite{99ADKLPS}. In that reference one can find
details of another data set that was analyzed in the same fashion, leading to
results in agreement with those reviewed here.

Firstly we tested the isotropy of the flow for separations of the order of $
\Delta $. Using the standard Taylor hypothesis, define the ``transverse''
structure function across $\Delta $ as $S_{T}(\Delta )\equiv \langle \lbrack
u_{1}(\bar{U}t)-u_{2}(\bar{U}t)]^{2}\rangle $ and the ``longitudinal''
structure function as $S_{L}(\Delta )\equiv \langle \lbrack u_{1}(\bar{U}t+
\bar{U}t_{\Delta })-u_{1}(\bar{U}t)]^{2}\rangle $ where $t_{\Delta }=\Delta /
\bar{U}$. If the flow were isotropic we would expect \cite{MY} 
\begin{equation}
S_{T}(\Delta )=S_{L}(\Delta )+{\frac{\Delta }{2}}{\frac{\partial
S_{L}(\Delta )}{\partial \Delta }}\ .  \label{Stl}
\end{equation}
In the isotropic state both components scale with the same exponent, $
S_{T,L}(\Delta )\propto \Delta ^{\zeta _{2}}$, and their ratio is computed
from ( \ref{Stl}) to be $1+\zeta _{2}/2\approx 1.35$ where $\zeta
_{2}\approx 0.69$ (see below). The experimental ratio was found 
to be 1.32, indicating that the anisotropy at the scale 
$\Delta$ is small. We expect that the effects of anisotropy should be 
most pronounced on the larger scales.

Next, we found the functional form of the basis tensors $B_{qjm}^{\alpha
\beta }({\bf \hat{R}})$ and the algebraic relations among the coefficients $
a_{qjm}$ according to the discussion in the last paragraph of Appendix \ref
{sec:general-form}.

Since the anisotropies are not huge, we focused on the lowest order
corrections to the isotropic ($j=0$) contribution. In other words, we wrote 
\begin{equation}
S^{\alpha \beta }({\bf R})=S_{j=0}^{\alpha \beta }({\bf R})+S_{j=1}^{\alpha
\beta }({\bf R})+S_{j=2}^{\alpha \beta }({\bf R})+S_{j=3}^{\alpha \beta }(
{\bf R}) \ .
\end{equation}
We defined the coordinate system such that the mean wind direction was along
the 3-axis, and the separation between the two probes was along the 1-axis.
By {\em assuming} axial symmetry along the mean wind direction, the tensors $
S_{j=0}^{\alpha \beta }({\bf R}),\;S_{j=1}^{\alpha \beta }({\bf R}
),\;S_{j=2}^{\alpha \beta }({\bf R})$ were to contain {\em only} the $m=0$
components. In addition, since the two probes measured the velocity field
only in the mean wind direction, we had only the values of $S^{33}({\bf R})$
in the $1-3$ plane. In such a case, it turns out that only the even $j$'s have a
non-vanishing contribution. We therefore used the trial tensor: 
\[
S^{\alpha \beta }({\bf R})=S_{j=0}^{\alpha \beta }({\bf R})+S_{j=2}^{\alpha
\beta }({\bf R}) \ , 
\]
\begin{eqnarray*}
S_{j=0}^{\alpha \beta }({\bf R}) &=&c_{0}\left( \frac{R}{\Delta }\right)
^{\zeta _{2}}\left[ (2+\zeta _{2})\delta ^{\alpha \beta }-\zeta _{2}{\frac{
R^{\alpha }R^{\beta }}{R^{2}}}\right] \ , \\
S_{j=2}^{\alpha \beta }({\bf R}) &=&aS_{j=2,q=1}^{\alpha \beta }({\bf R}
)+bS_{j=2,q=2}^{\alpha \beta }({\bf R}).
\end{eqnarray*}
Where $S_{j=2,q=1}^{\alpha \beta }({\bf R}),\;S_{j=2,q=2}^{\alpha \beta }(
{\bf R})$ are given by:


\begin{eqnarray}
S_{j=2,q=1}^{\alpha \beta }({\bf R}) &=&\left( \frac{R}{\Delta }\right)
^{\zeta _{2}^{(2)}}\Big[(\zeta _{2}^{(2)}-2)\delta ^{\alpha \beta }-\zeta
_{2}^{(2)}(\zeta _{2}^{(2)}+6)\delta ^{\alpha \beta }{\frac{(\B.k\cdot \B .
R)^{2}}{R^{2}}}+2\zeta _{2}^{(2)}(\zeta _{2}^{(2)}-2){\frac{R^{\alpha
}R^{\beta }(\B.k\cdot \B.R)^{2}}{R^{4}}}  \nonumber \\
&+&([\zeta _{2}^{(2)}]^{2}+3\zeta _{2}^{(2)}+6)k^{\alpha }k^{\beta }-{\frac{
\zeta _{2}^{(2)}(\zeta _{2}^{(2)}-2)}{R^{2}}}(R^{\alpha }k^{\beta }+R^{\beta
}k^{\alpha })(\B.k\cdot \B.R)\Big]  \label{finalform} \ , \\
S_{j=2,q=2}^{\alpha \beta }({\bf R}) &=&\left( \frac{R}{\Delta }\right)^{\zeta_2^{(2)}} \Big
[-(\zeta _{2}^{(2)}+3)(\zeta _{2}^{(2)}+2)\delta ^{\alpha \beta }(\B.k\cdot 
\B.R)^{2}+(\zeta _{2}^{(2)}-2){\frac{R^{\alpha }R^{\beta }}{R^{2}}} 
\nonumber \\
&+&(\zeta _{2}^{(2)}+3)(\zeta _{2}^{(2)}+2)k^{\alpha }k^{\beta }+(2\zeta
_{2}^{(2)}+1)(\zeta _{2}^{(2)}-2)  \nonumber \\
&\times &{\frac{{R^{\alpha }}{R^{\beta }}{(\B.k\cdot \B.R)^{2}}}{R^{4}}}
-([\zeta _{2}^{(2)}]^{2}-4)(R^{\alpha }k^{\beta }+R^{\beta }k^{\alpha })(\B .
k\cdot \B.R)\Big]\ .  \nonumber
\end{eqnarray}

The vector ${\B.k}$ was taken to be along the mean wind direction. $\zeta
_{2}$ is the isotropic exponent, while $\zeta _{2}^{(2)}$ is the $j=2$
exponent. $c_{0},\;a,\;b$ are the non-universal weights of the components of 
$S^{\alpha \beta }({\bf R})$. In order to reduce the number of unknown
quantities, the exponent of the isotropic part of $S^{\alpha \beta }({\bf R}
) $ was assumed to be known: $\zeta _{2}=0.69$.$\;$The values of $
c_{0},\;a,\;b,\;\zeta _{2}^{(2)}\;$were to be found from the experimental
data.

Using spherical coordinates, the trial tensor $S^{33}$ in the $1-3$ plane
took the following form: 
\begin{eqnarray}
S^{33}(R,\theta ,\phi &=&0)=S_{j=0}^{33}(R,\theta ,\phi
=0)+S_{j=2}^{33}(R,\theta ,\phi =0) \\
&=&c_{0}\left( {\frac{R}{\Delta }}\right) ^{0.69}\Big[2+0.69-0.69\cos
^{2}\theta \Big]  \nonumber \\
&&+a\left( {\frac{R}{\Delta }}\right) ^{\zeta _{2}^{(2)}}\Big[(\zeta
_{2}^{(2)}+2)^{2}-\zeta _{2}^{(2)}(3\zeta _{2}^{(2)}+2)\cos ^{2}\theta 
\nonumber \\
&&+2\zeta _{2}^{(2)}(\zeta _{2}^{(2)}-2)\cos ^{4}\theta \Big]  \nonumber \\
&&+b\left( {\frac{R}{\Delta }}\right) ^{\zeta _{2}^{(2)}}\Big[(\zeta
_{2}^{(2)}+2)(\zeta _{2}^{(2)}+3)-\zeta _{2}^{(2)}(3\zeta _{2}^{(2)}+4)\cos
^{2}\theta  \nonumber \\
&&+(2\zeta _{2}^{(2)}+1)(\zeta _{2}^{(2)}-2)\cos ^{4}\theta \Big].  \nonumber
\end{eqnarray}
where $\theta $ is the angle between ${\bf k}$ and ${\bf R}$. The fitting of
the trial tensor to the data was done along two paths in the $(R,\theta)$
space:

\begin{itemize}
\item  $\theta =0$ \ A single probe measurement.

\item  $R\sin \theta =\Delta $ \ Two probes measurement.
\end{itemize}

Fig. 1 shows the best fit to the data. For each type of data, two fits were
performed: A fit of the isotropic part only (panel a), and a fit of both
isotropic and $j=2$ components (panel b). The excellent fits in panel
(b) is a good support for the present mode of analysis.

In Ref.\cite{99ADKLPS} Fig.3 we showed the determination of $\zeta _{2}^{(2)}$ from a least-square
fit. The optimal value of this exponent and the uncertainty determined from
that plot is $\zeta _{2}^{(2)}=1.38\pm 0.15$. It should be understood that
the exponent $\zeta _{2}^{(2)}$ (and also $\zeta _{2}^{(1)}$ that is
unavailable from the present measurements) are just the smallest exponents
in the hierarchy $\zeta _{2}^{(j)}$ that characterizes higher order
irreducible representations indexed by $j$. The study of these exponents 
has only begun here, and considerable experimental and theoretical effort is
needed to reach firm conclusions regarding their universality and numerical
values. We expect the exponents to be a non-decreasing function of $j$,
explaining why the highest values of $j$ are being peeled off quickly when $
R $ decreases. Nevertheless, the lower order values of $\zeta _{2}^{(j)}$
can be measured and computed. In Ref.\cite{99ADKLPS} we presented an additional set
of experimental data, and demonstrated that the numerical value of 
$\zeta _{2}^{(2)}$ appears universal.

\section{Concluding Remarks}

The aim of this paper has been to introduce the exploration of the scaling
properties of turbulent statistics in the anisotropic sectors. The main
novel theoretical development is described in Sect.2. It is explained there
that the linearity of the equations for the fully unfused correlation
functions together with the invariance to rotations, foliates the solutions
into sectors characterized by the $j,m$ designations of the irreducible
representations of the SO(3) symmetry group. As a consequence we expect the
different sectors to be characterized by different (anomalous) scaling
exponents. This observation opens up a new and interesting research arena
for theory and experiments.

In Sect.3 we presented a derivation of the form of the second order
structure function in the higher $j$-sectors, and used a recent experimental
data analysis as an example of the utility and importance of the present
approach. The main result of this section, besides the theoretical forms
that can be used for further data analysis, is that the scaling range in
turbulence is much larger than expected. One just needs to acknowledge the
existence of anisotropic contributions to obtain scaling ranges that go all
the way from the Kolmogorov scale to the outer scale of turbulence.

It is our belief that this paper does not exhaust the issue of anisotropic
contributions to turbulent statistics. It is only the beginning of a rich
research program that should be carried simultaneously by experimentalist
and theorists. 
\acknowledgments
We are indebted to K.R. Sreenivasan and his group, notably Susan Kurien
and Brindesh Druvah, for the indispensable collaboration that resulted
in \cite{99ADKLPS} and the results mentioned in Sect.5.  
This work was supported in part by the Basic Research Fund administered by
the Israeli Academy of Sciences, he special Yale-Weizmann collaboration fund,
The German Israeli Foundation and the Naftali and Anna
Backenroth-Bronicki Fund for Research in Chaos and Complexity. 
\appendix
\section{The general form of the 2nd rank tensor}

\label{sec:general-form}

In this appendix we discuss the general structure of the second rank
correlation functions 
\begin{equation}
F^{\alpha \beta }({\bf R})\equiv\langle u^{\alpha }({\bf r}+{\bf R})u^{\beta }({\bf 
r})\rangle \ ,  \label{cf}
\end{equation}
In (\ref{cf}) {\em homogeneity} of the flow is assumed, but not isotropy.
Note that this object is more general than the structure function $S^{\alpha
\beta }$ in being nonsymmetric in the indices, and having no definite
parity. In light of the discussion in Sect. 2, when we expand this
objects in terms of tensors with definite $j,m$, we expect each component to
have a distinct dependence on the amplitude $R\equiv |{\bf R|}$.
Accordingly, we wish to find the basis functions $B_{qjm}^{\alpha \beta }( 
{\bf \hat{R}})$, with which we can represent $F^{\alpha \beta }({\bf R})$ in
the form: 
\begin{equation}
F^{\alpha \beta }({\bf R})=\sum_{qjm}a_{qjm}(R)B_{qjm}^{\alpha \beta }({\bf 
\hat{R}})  \label{trial-tensor}
\end{equation}
and derive some constraints among the functions $\ a_{qjm}(R)$ that result
from incompressibility. We shall see, that due to the isotropy of the
incompressibility conditions, the constraints are among $a_{qjm}(R)$ with
the {\em same} $j,m$ only.

We begin by analyzing the incompressibility condition: An incompressible
flow with constant density is characterized by the relation: 
\[
\partial _{\alpha }u^{\alpha }({\bf r},t)=0 
\]
as a result, one immediately gets the following constraints on $F^{\alpha
\beta }({\bf R})$: 
\begin{eqnarray*}
\partial _{\alpha }F^{\alpha \beta }({\bf R}) &=&0 \\
\partial _{\beta }F^{\alpha \beta }({\bf R}) &=&0.
\end{eqnarray*}
Plugging the trial tensor (\ref{trial-tensor}) into the last two equations
we obtain 2 equations connecting the different $a_{qjm}$: 
\begin{eqnarray}
\partial _{\alpha }\sum_{qjm}a_{qjm}(R)B_{qjm}^{\alpha \beta }({\bf \hat{R}}
) &=&0  \label{incomp-1} \\
\partial _{\beta }\sum_{qjm}a_{qjm}(R)B_{qjm}^{\alpha \beta }({\bf \hat{R}})
&=&0  \nonumber
\end{eqnarray}

We first notice that the differentiation action is isotropic. As a result,
if $T^{\alpha \beta }\left( {\bf R}\right) $ is some arbitrary tensor with a
definite $j,m$ transformation properties, then the tensor $\partial _{\alpha
}T^{\alpha \beta }({\bf R})$ will have {\em the same} $j,m$ transformation
properties. Components with different $j,m$ are linearly independent.
Therefore equations (\ref{incomp-1}) should hold for each $j,m$ separately.

Next, we observe that (\ref{incomp-1}) are invariant under the transformation $
F^{\alpha \beta }\longrightarrow F^{\beta \alpha }$. As a result, the
symmetric and anti-symmetric parts of $F^{\alpha \beta }$ should satisfy 
(\ref{incomp-1}) independently. To see that, let us write $F^{\alpha \beta }$
as a sum of a symmetric term and an anti-symmetric term: $F^{\alpha \beta
}=F_{S}^{\alpha \beta }+F_{A}^{\alpha \beta }$, we then get: 
\begin{eqnarray*}
\partial _{\alpha }F^{\alpha \beta } &=&\partial _{\alpha }F_{S}^{\alpha
\beta }+\partial _{\alpha }F_{A}^{\alpha \beta }=\partial _{\alpha
}F_{S}^{\beta \alpha }-\partial _{\alpha }F_{A}^{\beta \alpha }=0 \\
\partial _{\beta }F^{\alpha \beta } &=&\partial _{\beta }F_{S}^{\alpha \beta
}+\partial _{\beta }F_{A}^{\alpha \beta }=0
\end{eqnarray*}
from which we conclude: 
\[
\partial _{\alpha }F_{S}^{\alpha \beta }=\partial _{\alpha }F_{A}^{\alpha
\beta }=0 
\]

Finally, (\ref{incomp-1}) is invariant under the transformation $F^{\alpha
\beta }({\bf R)}\longrightarrow F^{\alpha \beta }(-{\bf R)}$ and as a result
the odd parity and the even parity parts of $F^{\alpha \beta }$ should
fulfill (\ref{incomp-1}) independently. We conclude that a necessary and
sufficient condition for (\ref{incomp-1}) to hold is that it holds
separately for parts with definite $j,m$, definite symmetry in the $\alpha
,\beta $ indices and a definite parity in ${\bf R}$: 
\[
\partial _{\alpha }\sum_{q}a_{qjm}(|{\bf R|})B_{qjm}^{\alpha \beta }({\bf 
\hat{R}})=0\text{ \ 
\begin{tabular}{l}
~~{\small summation is over} \\ 
~~$B_{qjm}^{\alpha \beta }$ {\small with definite symmetries}
\end{tabular}
} 
\]
where the summation is over $q$ such that $B_{qjm}^{\alpha \beta }$ has a
definite indices symmetry and a definite parity.

According to (\ref{eq:second-rank-tensors}) we can write these 
$B_{qjm}^{\alpha \beta }$ as:

\begin{enumerate}
\item  $(-)^{j}$ parity, symmetric tensors:

\begin{itemize}
\item  $B_{1,jm}^{\alpha \beta }(\hat{{\bf R}})\equiv R^{-j}\delta ^{\alpha
\beta }\Phi _{jm}(${\bf $R$}$)$,

\item  $B_{7,jm}^{\alpha \beta }(\hat{{\bf R}})\equiv R^{-j}\left[ R^{\alpha
}\partial ^{\beta }+R^{\beta }\partial ^{\alpha }\right] \Phi _{jm}(${\bf $R$
}$)$,

\item  $B_{9,jm}^{\alpha \beta }(\hat{{\bf R}})\equiv R^{-j-2}R^{\alpha
}R^{\beta }\Phi _{jm}(${\bf $R$}$)$,

\item  $B_{5,jm}^{\alpha \beta }(\hat{{\bf R}})\equiv R^{-j+2}\partial
^{\alpha }\partial ^{\beta }\Phi _{jm}(${\bf $R$}$)$.
\end{itemize}

\item  $(-)^{j}$ parity, anti-symmetric tensors:

\begin{itemize}
\item  $B_{3,jm}^{\alpha \beta }(\hat{{\bf R}})\equiv R^{-j}\left[ R^{\alpha
}\partial ^{\beta }-R^{\beta }\partial ^{\alpha }\right] \Phi _{jm}(${\bf $R$
}$)$.
\end{itemize}

\item  $(-)^{j+1}$ parity, symmetric tensors:

\begin{itemize}
\item  $B_{8,jm}^{\alpha \beta }(\hat{{\bf R}})\equiv R^{-j-1}\left[
R^{\alpha }\epsilon ^{\beta \mu \nu }R_{\mu }\partial _{\nu }+R^{\beta
}\epsilon ^{\alpha \mu \nu }R_{\mu }\partial _{\nu }\right] \Phi _{jm}(${\bf 
$R$}$)$,

\item  $B_{6,jm}^{\alpha \beta }(\hat{{\bf R}})\equiv R^{-j+1}\left[
\epsilon ^{\beta \mu \nu }R_{\mu }\partial _{\nu }\partial ^{\alpha
}+\epsilon ^{\alpha \mu \nu }R_{\mu }\partial _{\nu }\partial ^{\beta }
\right] \Phi _{jm}(${\bf $R$}$)$.
\end{itemize}

\item  $(-)^{j+1}$ parity, anti-symmetric tensors:

\begin{itemize}
\item  $B_{4,jm}^{\alpha \beta }(\hat{{\bf R}})\equiv R^{-j-1}\epsilon
^{\alpha \beta \mu }R_{\mu }\Phi _{jm}(${\bf $R$}$)$,

\item  $B_{2,jm}^{\alpha \beta }(\hat{{\bf R}})\equiv R^{-j+1}\epsilon
^{\alpha \beta \mu }\partial _{\mu }\Phi _{jm}(${\bf $R$}$)$.
\end{itemize}
\end{enumerate}

In order to differentiate these expressions we can use the following
identities: 
\begin{eqnarray*}
R^{\alpha }\partial _{\alpha }R^{\zeta }Y_{jm}(\hat{{\bf R}}) &=&\zeta
R^{\zeta }Y_{jm}(\hat{{\bf R}}) \ ,\\
\partial ^{\alpha }\partial _{\alpha }R^{\zeta }Y_{jm}(\hat{{\bf R}}) &=&
\left[ \zeta \left( \zeta +1\right) -j(j+1)\right] R^{\zeta -2}Y_{jm}(
\hat{x})
\end{eqnarray*}
which give rise to: 
\begin{eqnarray*}
R^{\alpha }\partial _{\alpha }\Phi _{jm}({\bf R}) &=&j\Phi _{jm}({\bf R}) \ .\\
\partial ^{\alpha }\partial _{\alpha }\Phi _{jm}({\bf R}) &=&0 \ .
\end{eqnarray*}
\ From this point, it is a matter of simple (though somewhat lengthy)
algebra to derive the differential constraints among $a_{qjm}(R)$. The
results are as follows:

\begin{enumerate}
\item  $q\in \{1,7,9,5\}$ 
\begin{eqnarray}
a_{1,jm}^{\prime }(R)-jR^{-1}a_{1,jm}+ja_{7,jm}^{\prime
}-j^{2}R^{-1}a_{7,jm}+a_{9,jm}^{\prime }+2R^{-1}a_{9,jm} &=&0 \ ,
\label{eq:in1795} \\
R^{-1}a_{1,jm}+a_{7,jm}^{\prime }+3R^{-1}a_{7,jm}+\left( j-1\right)
a_{5,jm}^{\prime }-\left( j^{2}-3j+2\right) R^{-1}a_{5,jm} &=&0  \ . \nonumber
\end{eqnarray}

\item  $q\in \{3\}$ 
\begin{eqnarray}
a_{3,jm}^{\prime }-jR^{-1}a_{3,jm} &=&0  \label{eq:in3} \ , \\
a_{3,jm}^{\prime }+R^{-1}a_{3,jm} &=&0 \ . \nonumber
\end{eqnarray}

{\bf Notice}: These equations have no solutions other than: $a_{3,jm}(R)=0$.

\item  $q\in \{8,6\}$ 
\begin{equation}
a_{8,jm}^{\prime }+3R^{-1}a_{8,jm}+(j-1)a_{6,jm}^{\prime }-\left(
j^{2}-2j+1\right) R^{-1}a_{6,jm}=0  \ . \label{eq:in86}
\end{equation}

\item  $q\in \{4,2\}$ 
\begin{equation}
R^{-1}a_{4,jm}-a_{2,jm}^{\prime }+(j-1)R^{-1}a_{2,jm}=0 \ . \label{eq:in42}
\end{equation}
\end{enumerate}

There are obviously more unknowns than equations, since we merely exploited
the incompressibility conditions. Nevertheless, we believe
that the missing equations that arise from the dynamical hierarchy of
equations will preserve the distinction between $a_{qjm}$ of different $j,m$
(again, due to the isotropy of these equations).

Note also, that the above analysis holds also for the second-order structure
function 
\[
S^{\alpha \beta }({\bf R})\equiv \left\langle \left[ u^{\alpha }({\bf r}+
{\bf R})-u^{\alpha }({\bf r})\right] \left[ u^{\beta }({\bf r}+{\bf R}
)-u^{\beta }({\bf r})\right] \right\rangle . 
\]
Only that in this case we should only consider the representations $
q=1,7,9,5 $ for even $j$ and the representations $q=8,6$ for odd $j$. This
follows from the fact that $S^{\alpha \beta }({\bf R})$ is symmetric with
respect to its indices and it has an even parity in \ ${\bf R}$. Also, in
that case, it is possible to go one step further by assuming a specific
functional form for the $a_{q,jm}(R)$. We know that the $S^{\alpha \beta }(
{\bf R})$ is expected scale in the inertial range, and we therefore may {\em 
assume}: 
\[
a_{q,jm}(R)\equiv c_{q,jm}R^{\zeta _{2}^{(j)}}. 
\]
where $c_{q,jm}$ are just numerical constants. If we now substitute this
definition into the equations (\ref{eq:in1795},\ref{eq:in86}), we get a set
of linear equations among the $c_{q,jm}$. These relations can be easily
solved and give us two possible tensors for even $j$ ($q=1,7,9,5$) and one
tensor form for odd $j$ (from $q=8,6$). This kind of approach was taken in
the two-probes experiment which is described in Sect. \ref{sec:experiment}.

\newpage
\begin{figure}
\epsfxsize=8.5truecm
\epsfbox{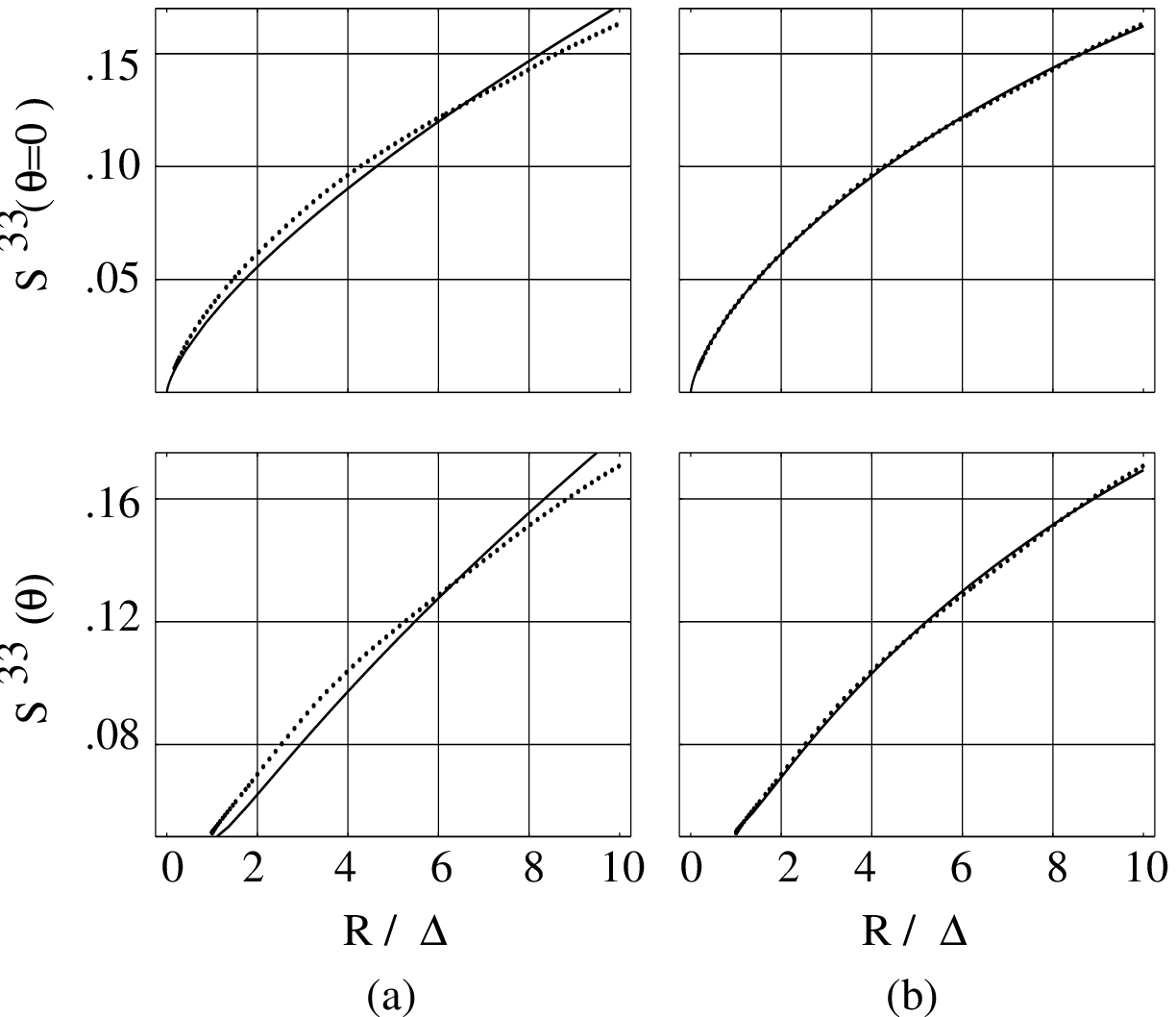}
\caption{The structure functions $S^{33}$ for $\theta=0$ and for 
non-zero $\theta$ computed for set I. The dots are for experimental 
data and the line is the analytic fit. Panel (a) presents fits to the 
$j=0$ component only, and panel (b) to components $j=0$ and $j=2$ together.} 
\label{Fig.1} \end{figure}
\end{document}